\documentclass[aps,showpacs,amsmath
]{revtex4}

\usepackage[dvips]{graphicx}
\usepackage{bm}
\usepackage{longtable}
\usepackage{amsmath}
\usepackage{dcolumn}
\usepackage{placeins}
\usepackage{amsbsy}

\begin{document}

\bibliographystyle{apsrev}

\title {Colour-octet bound states, induced by Higgs mechanism}

\author{S. Bladwell$^1$} \email{samuel.s.bladwell@unsw.edu.au}
\author{V. F. Dmitriev$^2$}
\author{V. V. Flambaum$^1$} \email[Email:]{flambaum@phys.unsw.edu.au}
\author{A. Kozlov$^1$} \email[Email:]{a.kozloff@unsw.edu.au}
\affiliation{$^1$School of Physics, University of New South Wales, Sydney, 2052, Australia.}
\affiliation{$^2$Budker Institute for Nuclear Physics and Novosibirsk State University, Novosibirsk, 630090, Russia.}

    \date{\today}

    \begin{abstract}

The current limits for fourth generation quarks allows to expect their mass of the order of 500 GeV. In this mass region for quark-anti-quark pair the additional Yukawa-type attraction due to Higgs mechanism is expected to emerge. This Higgs induced attraction greatly exceeds strong interaction between quarks and leads to the formation of bound states in both colour octet $S^{(8)}$ and singlet $S^{(1)}$ states. In the key of recent works on significance of colour octet channel for production of colour singlet state of fourth generation $Q\overline{Q}$ we calculated the binding energies for both octet and singlet states. Such attraction localizes quarks in extremely small area. Hence colour octet pair of fourth generation quarks can form the "nucleus" and together with colour neutralizing light particle that is captured by strong interaction in orbit around the nucleus, create particle, similar by its structure to Deuterium.
    \end{abstract}


\maketitle

Production of quarkonium includes both colour-singlet and colour-octet contributions \cite{Cho:1996}. The octet state can decay into singlet via soft (low energy) gluon emission \cite{Wong:1999}. As was shown in \cite{Bodwin:1995, Braaten:1999, Braaten:1995, Leibovich:2001} in certain cases colour-octet mechanism can make dominant contribution to quarkonia production crossection. Since Higgs coupling strength is proportional to the fermion masses the influence of Higgs field on colour-octet state of heavy 4G (4th generation) $Q\overline{Q}$ can be significant.

As was discussed in \cite{Enkhbat:2011} Yukawa-type attraction can lead to the appearance of colour octet 4G $Q\overline{Q}$ ground state. Authors considered pseudo-scalar Nambu-Goldstone mechanism as the main source of coupling. But the latter mechanism requires relativistic effects to be significant while scalar Higgs exchange leads to significant effects already in non-relativistic region. Another model with two Higgs doublets was discussed in \cite{GAKozlov:2004}. In the light of recent CERN report on discovery of new particle with some properties expected for Higgs boson it appears reasonable to consider it as a main mechanism for 4G quarkonium coupling source. In \cite{Flambaum:2011} there was shown that such mechanism significantly increases binding energy of colour-singlet state of 4G quarkonium. In colour-octet state Higgs induced attraction can be the only source of formation of bound state since gluon exchange creates effective repulsion in this case. To show this it is convenient to use variational approach, suggested in \cite{Flambaum:2011}. The Hamiltonian of the system is given by the following equation:
\begin{equation}
H_{0}=T + V_\text{Y}(r)+V_\text{Strong}(r),
\end{equation}
where $T$ is relativistic kinetic energy, $V_\text{Y}$ is Yukava-type interaction induced by Higgs exchange:
\begin{equation}
V_Y(r)=-\frac{\alpha_h}{r}\exp(-m_h r).
\end{equation}
Here $m_h$ is Higgs mass, $\alpha_h=m^2/(4\pi v^2)$ is the Higgs field coupling constant. According to standard model Higgs vacuum expectation value $v=246$ GeV.
Coulomb-like  strong interaction $V_\text{Strong}$ is given by the following expression:
\begin{equation}
V_\text{Strong}=-\frac{a}{r},
\end{equation}
where constant $a=-\alpha_s/6$ for octet state and $a=4\alpha_s/3$ for singlet ($\alpha_s$ is strong coupling constant). As one can see, for octet state strong interaction creates effective repulsion.

The total energy of the system can be written as
\begin{equation}
\tilde E = \left< T \right>  + \left< V_\text{rel} \right> +  \left< V_\text{rad} \right> + \left< V_\text{Y} \right>  +  \left< V_\text{Strong} \right>,
\label{octetenergy}
\end{equation}
where $ \left< V_\text{rel} \right>$ and  $\left<V_\text{rad}\right>$ represents relativistic and radiative corrections respectively. The variational ground-state wave function is taken in hydrogen like form:
\begin{equation}
\psi(r)=\pi^{-1/2}q^{3/2}\exp(-qr).
\label{wave}
\end{equation}
The exact expressions of all the operators in (\ref{octetenergy}) can be found in \cite{Flambaum:2011}. We will present the answers for their values averaged over wave function (\ref{wave}):
\begin{align}
&\frac{\left< T \right>}{m}=\frac{4}{\pi}\left\{\frac{x(3-4x^2+4x^4}{3(1-x^2)^2}+\frac{(1-2x^2)\arccos(x)}{(1-x^2)^{5/2}}\right\}\\
&\left< V_\text{Y} \right>=-\frac{4\alpha_h q^3}{(m_h+2q)^2}\\
&\left< V_\text{Strong} \right>=-aq\\
&\left< V_\text{rel} \right>=\frac{6\alpha_h q^4}{m^2 (2q+m_h)}\\
&\left< V_\text{rad} \right>=\frac{4\alpha_h^2q^3}{\pi m^2}\left(\gamma - \frac{\nu}{20}-\frac{4\pi\gamma m_h^2}{(m_h+2q)^2}\right),
\end{align}
where $\nu=11$ for 4G quark - the number of heavy fermions in polarization loop and coefficient $\gamma$ is given by the following expression:
\begin{equation}
\gamma=\frac{1}{3}\left(\ln\frac{m+m_h}{m_h}-\frac{7m}{4m+5m_h}\right).
\end{equation}
As soon as the expression (\ref{octetenergy}) for the ground-state energy $E=\min\{\tilde E(q)\}$ is defined it should be minimized over parameter $q$. With a precision of several percent expression (\ref{octetenergy}) reaches it's minimum at
\begin{equation}
q_{min}=\frac{m}{9(\tilde\alpha_h+a)}(\sqrt{1+9(\tilde\alpha_h+a)^2}-1),\label{qmin}
\end{equation}
where
\begin{equation}
\tilde\alpha_h=\alpha_h\left(1+3\frac{m_h^2}{m^2(\alpha_h+a)^2}\right)^{-1}.
\end{equation}

\begin{figure}
\includegraphics[width=0.45\textwidth]{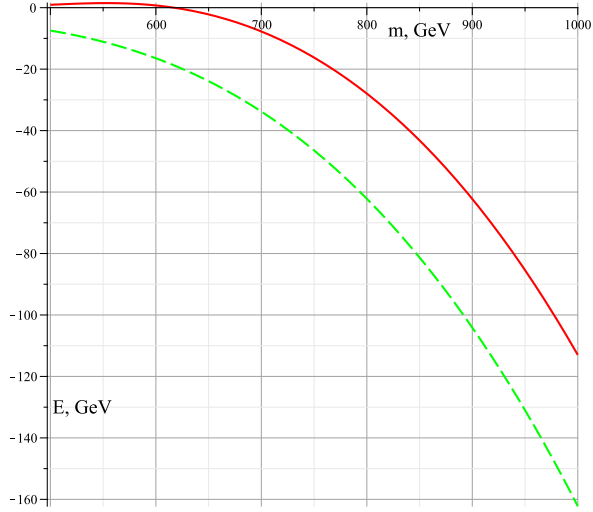}
\caption{The dependence of ground state energy $E$ on the particles mass $m$. Solid and dashed lines represents colour-octet and colour-singlet states respectively.}\label{Fig:1}
\end{figure}

The results for binding energy $E$ are presented at Fig.~\ref{Fig:1}. As one can see, binding energy always exists for singlet state, as was expected, since both Higgs induces and strong interactions are attractive. For octet ground state the situation is different. Up to the masses $m < m_{cr}\approx 620$ GeV repulsive strong interaction dominates over Higgs attractive one. But for $m > m_{cr}$ the ground state energy becomes negative, therefore for quark masses $m>620$ GeV it is reasonable to expect existence of bound states. But since octet state itself has larger energy compared to singlet one it should decay over time. Besides such a heavy 4G quarks itself should experience weak decay to lighter quarks as well as other decay channels \cite{Enkhbat:2011}. Since in this paper we are mainly interested on octet channel contribution to production of colour singlet quarkonium below we will calculate its decay width and briefly discuss other decay channels afterwards.

As can be concluded from Fig. \ref{Fig:1} for any mass of the fermions octet state have greater energy then singlet state. It leads to the decay of octet state into the singlet state with emission of gluon. This decay of octet state quarkonium contributes to the total production cross section of the singlet state quarkonium. To estimate the width of this transition we consider the emission of a gluon as M1 transition, analogous to M1 transition with photon emission, but with addition of relevant colour factor. This perturbative approach expected to be valid at least qualitatively, since the emitted gluon is hard enough, its energy is greater than 10 GeV  (see Fig.1), therefore we expect that nonperturbative corrections could be small due to their power dependence on gluon energy.

The decay width of M1 transition is given be the equation
\begin{equation}
\Gamma= 2\pi |V_{fi}|^2,
\label{M1}
\end{equation}
where $V_{fi}$ is the matrix element of M1 transition. The general formula for M1 amplitude of photon emission can be found in \cite{Berestetskii:1971}
\begin{equation}
V_{fi} = (-1)^m i \sqrt{\frac{6}{\pi}} \frac{\omega^{3/2}}{3} e ({Q_{1, -m}^{(m)}})_{fi},
\label{matrix}
\end{equation}
where $({Q_{j, -m}^{(m)}})_{fi}$ is so called 2$^j$ magnetic moment of transition, $\omega=\triangle E$ is frequency of emitted photon. For $^1S_0\rightarrow {}^3S_1$ M1 transition $j=1, l=0$ and in non-relativistic limit $e({Q_{j, -m}^{(m)}})_{fi}=\mu_{fi}$, where $\mu_{fi}$ is magnetic moment of transition:
\begin{equation}
\boldsymbol{\mu}_{fi}=\int \Psi_f^*\mu\frac{\mathbf s}{s}\Psi_i d^3r,
\label{mu}
\end{equation}
where $\mu$ and $s$ are magnetic moment of the particle and its spin. To perform transition from photon to gluon emission firstly it is needed to express the final and initial wave-functions in above integral as a product of spin, colour and coordinate parts:
\begin{equation}
\Psi_{i} = \left| ^1S_0 \right> \left| C^8\right> \psi_i \quad \Psi_{f} = \left| ^3S_1 \right> \left| C^1\right> \psi_f
\label{psi}
\end{equation}
Secondly, replace magnetic moment $\mu$ with chromomagnetic moment of the quark $\mu_{t'}\lambda_\rho/2$, where $\lambda_\rho$ are Gell-Mann matrices, carry out the summation over colour index $\rho=1..8$. The value of chromomagnetic moment $\mu_{t'}=-\mu_{\overline t'}=g/m_{t'}$ \cite{Wong:1999}, where $g$ is the colour charge. Substituting (\ref{psi}) and (\ref{mu}) to (\ref{matrix}) we obtain the expression for squared matrix element of transition
\begin{align}
&|V_{fi}|^2= \frac{2}{3\pi}\omega^3  \left| \int  \psi^*_f \psi_i d^3 r \right|^2\times\nonumber\\
&\overline{\left|\left< ^3S_1 \right|\mu \mathbf s\left| ^1S_0 \right>\right|^2} \sum_\rho\left|\left< C^1 \right| \frac{\lambda_\rho }{2} \left| C^8 \right> \right|^2,
\end{align}
where the averaging over total spin projections of initial state gives:
\begin{equation}
\overline{\left|\left< ^3S_1 \right|\mu \mathbf s\left| ^1S_0 \right>\right|^2} =4\frac{g^2}{m^2}.
\end{equation}
Fulfilling summation over colour states
\begin{equation}
\sum_{\rho}\left|\left< C^1\right| \frac{\lambda_\rho }{2} \left| C^8 \right> \right|^2 = \frac{1}{6}
\end{equation}
quantum numbers one can obtain the following expression for total decay width
\begin{equation}
\Gamma=\frac{4 g^2\omega^3}{9 m^2}\left| \int  \psi^*_f \psi_i d^3 r \right|^2.
\label{gamma81}
\end{equation}
The wave-functions $\psi_f$ and $\psi_i$ of singlet and octet states respectively can be estimated using (\ref{wave}) and the minimum values of parameters for singlet $q_{min}^{(1)}$ and octet $q_{min}^{(8)}$ are given by (\ref{qmin}) with corresponding values of constant $a$:
\begin{equation}
\left| \int  \psi^*_f \psi_i d^3 r \right|^2=64\frac{\left(q_{min}^{(1)}q_{min}^{(8)}\right)^3}{\left(q_{min}^{(1)}+q_{min}^{(8)}\right)^6}.
\end{equation}
After substituting above expression to (\ref{gamma81}) the final expression for the octet-singlet decay width is given by the following expression:
\begin{equation}
\Gamma=\frac{256g^2}{9}\frac{\omega^3}{m_{t'}^2}\frac{\left(q_{min}^{(1)}q_{min}^{(8)}\right)^3}{\left(q_{min}^{(1)}+q_{min}^{(8)}\right)^6}.
\label{decay81}
\end{equation}

\begin{figure}
\includegraphics[width=0.45\textwidth]{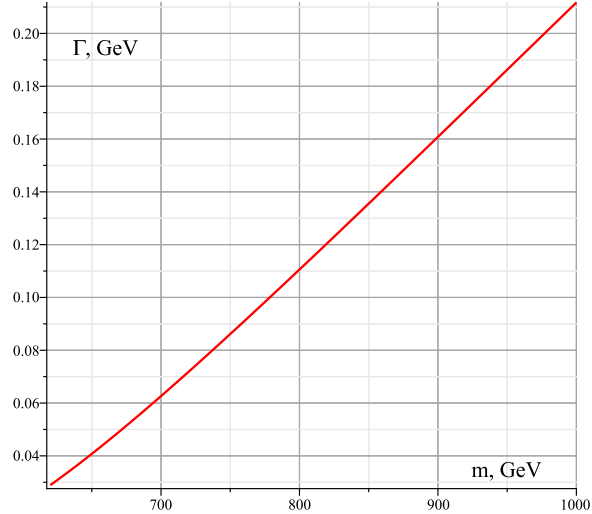}
\caption{The decay width (\ref{decay81}) of heavy quarkonium octet state $^3S_1^{(8)}$ to singlet state $^1S_0^{(1)}$ transition as a function of quark mass $m$.}\label{Fig:2}
\end{figure}

As can be seem from Fig.~\ref{Fig:2}, decay width of octet-singlet transition is almost two orders of magnitude less then the corresponding binding energies, shown on Fig.~\ref{Fig:1}. The rest of the significant decay channels were discussed and estimated in \cite{Enkhbat:2011}. All of those decay channels were shown to be small as well, except the semi-leptonic $t'\overline{t}'\rightarrow b'\overline{t'}\, W$. The latter one was shown to be of the order of 20 GeV for large (greater then $M_W$) $m(t') - m(b')$ mass difference which indeed can take place in assumption of ultra-heavy Higgs boson. In recent paper \cite{Kribs:2007} there was shown that the mass difference of 4G quarks should be estimated using the following expression to be in agreement with current experimental data

\begin{equation}
m_{t'} - m_{b'} \approx \left(1 + \frac{1}{5} \ln \left(\frac{m_h}{115\text{GeV}} \right)\right)\times 55\text{GeV}.
\label{masslimits}
\end{equation}

As one can see, putting the Higgs mass $m_h=125$ GeV the 4G mass difference $\triangle m=m_{t'} - m_{b'}$ becomes less then the mass of $W$ boson, therefore the semi-leptonic decay channel should be suppressed as was noted in \cite{Enkhbat:2011}.

The above results show that accounting for Higgs induced attraction leads to the emergence of a bound state. The energy of bound state is at least order of magnitude greater then the total decay width. Therefore this effect should lead to sharp maximum in 4G colour octet production cross-section.

The long-living colour-octet bound state of 4G quark pair has several orders smaller size compared to the known mesons. It immediately follows from the relative strength of strong interaction and Higgs induced attraction and heaviness of 4G quarks:
\begin{equation}
r_H\sim\frac{1}{m_{t'}\alpha_h}
\end{equation}
As a consequence such a pair may capture another colour-charged particle, that would be bound to 4G "nucleus" by strong attraction, just like an electron is captured by Coulumb interaction in Deuterium. The comparison with Deuterium is not an accident, the ratio of sizes of nucleus and electron shell is defined by the ratio of strong nuclear interaction and Coulumb interaction in the same way as considered above object would have "electron" attached to the "nucleus" with strong interaction, while the nucleus itself is created by Higgs induced attraction.

\section{aknowledgement}

This work was supported in part  (VFD) by the Ministry of Science and Education of the Russian Federation, the RFBR grant 12-02-91341-NNIO\_a, and by Australian Research Council. One of the authors (VFD) acknowledges support from UNSW Faculty of Science Visiting Research Fellowships and Godfrey fund.

\end{document}